\documentclass[aps,pre,preprint,groupedaddress]{revtex4}

\usepackage{graphicx}
\usepackage{latexsym}
\usepackage{amsmath,amssymb}

\begin{document}


\title{
Finite-size scaling of the $d=5$ Ising model 
embedded in the cylindrical geometry:
An influence of the hyperscaling violation
}

\author{Yoshihiro Nishiyama}
\affiliation{Department of Physics, Faculty of Science,
Okayama University, Okayama 700-8530, Japan}

\date{\today}

\begin{abstract}
Finite-size scaling (FSS) of the five-dimensional ($d=5$) Ising model
is investigated numerically.
Because of the hyperscaling violation in $d>4$,
FSS of the $d=5$ Ising model no longer obeys the conventional
scaling relation.
Rather, it is expected that
the FSS behavior depends on the geometry
of the embedding space
(boundary condition).
In this paper, we consider the cylindrical geometry,
and explore its influence on the correlation length
$\xi=L^\Omega f( \epsilon L^{y^*_t} ,H L^{y^*_h} )$
with
system size $L$, reduced temperature $\epsilon$,
and magnetic field $H$;
the indices, $y^*_{t,h}$, and $\Omega$, characterize FSS.
For that purpose,
we employed the transfer-matrix method with Novotny's 
technique,
which enables us to treat
an arbitrary (integral) number of spins $N=8,10, \dots ,28$;
note that conventionally, $N$ is restricted in
$N(=L^{d-1})=16,81,256,\dots$.
As a result, we estimate the scaling indices as $\Omega=1.40(15)$,
$y^*_t=2.8(2)$, and $y^*_h=4.3(1)$.
Additionally, under postulating $\Omega=4/3$, we arrive at
$y^*_t=2.67(10)$ and $y^*_h=4.0(2)$.
These indices differ from the
naively expected ones, $\Omega=1$, $y^*_t=2$ and $y^*_h=3$.
Rather, our data support the generic formulas,
$\Omega=(d-1)/3$, $y^*_t=2(d-1)/3$, and $y^*_h=d-1$,
advocated for the cylindrical geometry in $d \ge 4$.
\end{abstract}


\maketitle

\section{\label{section1}Introduction}

The criticality
of the Ising model above the upper critical dimension ($d>4$)
belongs to the mean-field universality class.
However, the finite-size effect, namely,
the finite-size-scaling behavior, is not quite universal
because of the violation of the hyperscaling in $d>4$.
(From a renormalization-group viewpoint, this peculiarity 
is attributed to the presence of ``dangerous irrelevant variable'' 
\cite{Vladimir83,Privman83,Privman84}.)
Actually, it is expected that
the embedding geometry of the system (boundary condition)
would affect
the finite-size-scaling behavior for $d>4$ 
\cite{Brezin82,Brezin85,Privman90,%
Binder85,Binder85b,Parisi96,Lai90,Kenna91}.

Recently, 
Jones and Young performed an
extensive Monte Carlo simulation for
the $d=5$ Ising model embedded in the (periodic) hypercubic geometry
\cite{Jones05}.
They calculated the
correlation length $\xi$ with Kim's technique \cite{Kim93}.
(Because the calculation of $\xi$ requires a computational effort,
Binder's cumulant ratio rather than $\xi$ has been studied extensively so far
\cite{Binder85,Binder85b,Parisi96}.)
They found that the correlation length $\xi$ obeys the scaling relation,
\begin{equation}
\label{scaling_for_xi}
\xi  = L^\Omega f(\epsilon L^{y^*_t})  ,
\end{equation}
with the linear dimension $L$ and the reduced temperature $\epsilon$.
They found that 
the scaling indices are in good agreement with the
theoretical prediction
\cite{Brezin82,Brezin85,Privman90},
\begin{equation}
\label{hypercubic}
\Omega =  5/4 \ and \  y^*_t  =  5/2   .
\end{equation}

Notably enough, their indices exclude the naively expected values,
$\Omega=1$ and $y^*_t=2$.
That is, the correlation length $\xi \sim L^\Omega$ 
exceeds the system size $L$ at the critical point $\epsilon=0$ as $L\to \infty$.
(In other words, the spin-wave excitation costs very little energy
for large $L$.)
Such a peculiarity should be attributed to 
the violation of the
conventional scaling relation (hyperscaling) in $d>4$.
It would be intriguing that the above formula is cast into 
the expression \cite{Binder85},
\begin{equation}
\xi = l f(\epsilon l^2)   ,
\end{equation}
with the replacements $l=L^\Omega$ and $y^*_t=2\Omega$.
The expression is now reminiscent of the conventional scaling 
relation expected for the mean-field universality class.
Namely, the violation of hyperscaling is reconciled (absorbed)
by the replacements, and
the scaling indices, $\Omega$
and $y^*_t$, characterize the anomaly quantitatively.
So far, numerous considerations have been made
\cite{Binder85,Binder85b,Parisi96,Jones05} 
for the hypercubic geometry, 
where the Monte Carlo method works very efficiently.

In this paper, we investigate the $d=5$ Ising model embedded in the cylindrical geometry;
namely, we consider a system with infinite system size along a particular
direction.
Clearly, the transfer-matrix method well suits exploiting such a geometry.
However, practically,
the transfer-matrix method does not apply very well
in large dimensions $d \ge 3$
because of its severe limitation
as to the available system sizes.

In order to resolve this limitation,
we implemented Novotny's technique
\cite{Novotny90,Novotny92,Novotny93,Novotny91,Nishiyama04},
which enables us to
treat an arbitrary number of system sizes $N=8,10, \dots, 28$; 
here, the system size $N$
denotes the number of constituent spins within a unit of the transfer matrix 
(Fig. \ref{figure1}).
Note that conventionally, 
the system size is restricted in $N(=L^{d-1})=16,81,256,\dots$,
which soon exceeds the limit of available computer resources.
Such an arbitrariness allows us to treat a variety of system sizes,
and manage systematic finite-size-scaling analysis.
Actually, with the scaling analysis,
we obtained the indices $\Omega=1.40(15)$, $y^*_t=2.8(2)$, and $y^*_h=4.3(1)$.
Moreover, postulating $\Omega=4/3$, we obtained
$y^*_t=2.67(10)$ and $y^*_h=4.0(2)$.
(Here, the exponent $y^*_h$ denotes the scaling dimension of the magnetic field.
Our scaling relation, Eq. (\ref{cylinder}), incorporates the magnetic field
$H$ and the corresponding scaling index $y^*_h$.)
Obviously, our results exclude the naively expected ones,
$\Omega=1$, $y^*_t=2$, and $y^*_h=3$.
Rather, our data seem to support the generic formulas 
\cite{Brezin82,Brezin85,Privman90},
\begin{equation}
\label{ippan_cylinder}
\Omega  =  \frac{d-1}{3} , \ 
y^*_t   =   2\frac{d-1}{3} , \ and \ 
y^*_h  = d-1  ,
\end{equation}
advocated for the cylindrical geometry in $d \ge 4$.
Actually,
our data deviate from the above mentioned values for the hypercubic geometry,
Eq. (\ref{hypercubic}), indicating that the embedding geometry is indeed influential 
upon the finite-size scaling.

In fairness, its has to be mentioned that 
Novotny obtained $\Omega+y^*_t \approx d-1$ \cite{Novotny90,Novotny91}.
He postulated $\Omega=(d-1)/3$ in order to fix
the location of the critical point.
In this paper, we do not rely on any propositions, and
estimate the indices independently.
For that purpose, we calculated the cumulant ratio
to get information on the critical point.
Moreover, we treated the system sizes up to $N=28$,
which is
substantially larger than that of Ref. \cite{Novotny91}, $N \le 13$.
Here, we made use of an equivalence between the $d=5$ Ising model
and the quantum $d=4$ Ising model;
the latter is computationally less demanding.
We also eliminated 
insystematic finite-size corrections
by tuning extended coupling constants; see our Hamiltonian (\ref{Hamiltonian}).
In this respect, the motivation of the present research is well directed to
methodology.

The rest of this paper is organized as follows.
In Sec. \ref{section2}, we explain our simulation scheme in detail.
In Sec. \ref{section3}, we manage the finite-size-scaling analyses
of the simulation data.
In the last section, we present the summary and discussions.

\section{\label{section2}Numerical method}

In this section, we explain the numerical method.
First, we argue the reduction of the $d=5$ Ising model
to the $d=4$ quantum transverse-field Ising model.
The reduced (quantum mechanical) model is much easier to treat
numerically.
Second, we explicate Novotny's transfer-matrix method.
We place an emphasis how
we extended his formalism to adopt the quantum-mechanical interaction.

\subsection{\label{section2_1}Reduction of the classical $d=5$ Ising model
to the $d=4$ quantum counterpart}

The $d$-dimensional Ising model reduces to
the $(d-1)$-dimensional transverse-field Ising model;
in general, the $d$-dimensional classical system 
has its $(d-1)$-dimensional quantum counterpart \cite{Suzuki76}.
Such a reduction is based on the observation that
the transfer-matrix direction and the (quantum) imaginary-time evolution
have a close relationship.
Actually,
the quantum Hamiltonian is an infinitesimal generator 
of the transfer matrix.
Because the quantum Hamiltonian contains few non-zero elements,
its diagonalization requires reduced computational effort.
The significant point is that the universality class
(criticality) is maintained through the mapping.

To be specific, we consider the following $d=4$
transverse-field Ising model with the extended interactions.
The Hamiltonian is given by,
\begin{equation}
\label{Hamiltonian}
{\cal H}=
  -J_1\sum_{\langle ij \rangle}  \sigma^z_i \sigma^z_j
-J_2 \sum_{\langle \langle ij \rangle \rangle} \sigma^z_i \sigma^z_j
-J_3 \sum_{[ij]} \sigma^z_i \sigma^z_j
-J_4 \sum_{[[ij]]} \sigma^z_i \sigma^z_j 
-\Gamma \sum_i \sigma^x_i 
-H \sum_i \sigma^z_i 
   .
\end{equation}
Here, the operators $\{ \sigma^\alpha_i \}$ denote the Pauli matrices placed at the
$d=4$ hypercubic lattice points $i$.
The parameters $\Gamma$ and $H$ stand for the transverse and longitudinal
magnetic fields, respectively.
The summations,
$\sum_{\langle ij\rangle}$,
$\sum_{\langle \langle ij\rangle\rangle}$,
$\sum_{[ij]}$,
and
$\sum_{[[ij]]}$,
run over all possible nearest-neighbor pairs,
the next-nearest-neighbor (plaquette diagonal) pairs, 
the third-neighbor pairs, and the fourth-neighbor pairs, respectively.
The parameters $\{ J_i \}$ ($i=1,2,3,4$) are the corresponding
coupling constants.
Hereafter, we regard $J_1$ as a unit of energy ($J_1=1$),
and tune the remaining coupling constants 
$J_{2,3,4}$ so as to eliminate
the insystematic finite-size errors; see Sec. \ref{section3}.

We simulate the above $d=4$ quantum Ising model 
with the numerical diagonalization method.
The diagonalization of such a high-dimensional system
requires huge computer-memory space.
In fact, the number of spins constituting the $d=4$ cluster 
increases very rapidly as $N=16,81,256,\dots$,
overwhelming the available computer resources.
In the next section, we resolve this difficulty 
through resorting to Novotny's transfer-matrix formalism.

\subsection{\label{section2_2}
Constructions of the Hamiltonian matrix elements}

In this section,
we present an explicit representation for the
Hamiltonian (\ref{Hamiltonian}).
We make use of Novotny's method \cite{Novotny90},
which enables us to treat
an arbitrary number of spins constituting a unit of
the transfer matrix.
Novotny formulated the idea for the classical Ising model.
In this paper, we show that his idea is applicable to
the quantum Ising model, Eq. 
(\ref{Hamiltonian}), as well.

\begin{figure}
\includegraphics[width=100mm]{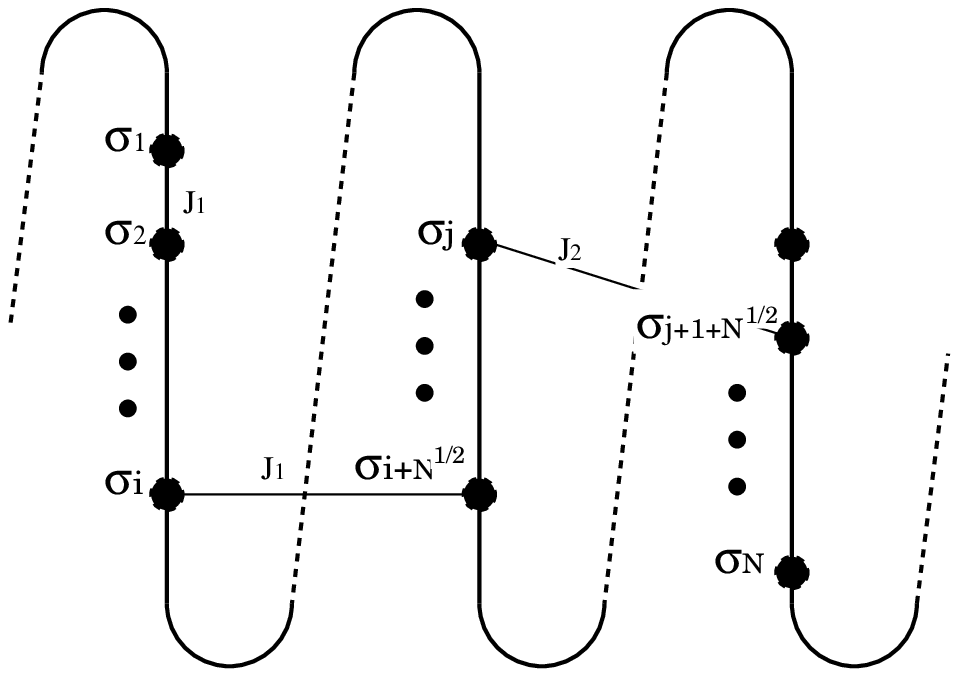}%
\caption{
\label{figure1}
Construction of the spin cluster for the quantum transverse-field Ising model, 
Eq. (\ref{Hamiltonian}).
For simplicity, we consider the case of $d=2$.
As indicated above, the spins constitute a $d=1$-dimensional alignment
$\{ \sigma_i \}$ ($i=1,2,\dots,N$),
and the dimensionality is lifted to $d=2$
by introducing the bridges
(long-range interactions)
over the $N^{1/2}$th and $1+N^{1/2}$th neighbor pairs;
these interactions correspond to the nearest-neighbor and the next-nearest-neighbor
interactions, respectively, with respect to the $d=2$ cluster.
In the case of $d=4$, we consider the $N^{1/4}$th, $N^{1/2}$th and $N^{3/4}$th
neighbor interactions; see Eq. (\ref{Hamiltonian2}) for detail.
This idea, namely, Novotny's method, was first developed for the classical Ising model
\cite{Novotny90}.
We apply this method to the quantum system.
}
\end{figure}

Before we commence a detailed discussion,
we explain the basic idea
of Novotny's method.
In Fig. \ref{figure1},
we presented a schematic drawing of a unit of the transfer matrix
for the Ising model in $d=3$ (rather than $d=5$ for the sake of simplicity).
Because the cross-section of the $d=3$-dimensional bar is $d=2$-dimensional,
the transfer-matrix unit should have a $d=2$-dimensional structure.
However, in Fig. \ref{figure1},
the spins $\{ \sigma_i \}$ 
($\sigma_i=\pm1$, $i=1,2, \dots ,N$) constitute a $d=1$-dimensional
(zig-zag) structure.
This feature is essential for us to construct the transfer-matrix unit 
with an arbitrary (integral) number of spins.
The dimensionality is lifted to $d=2$ effectively by the
long-range interactions over the $\sqrt{N}$th-neighbor pairs;
owing to the long-range interaction,
the $N$ spins constitute
a $\sqrt{N} \times \sqrt{N}$ rectangular network.
(The significant point is that the number $\sqrt{N}$ is not necessarily
an integral nor rational number.)
Similarly, the bridge over $(\sqrt{N} \pm 1)$th neighbor
pairs introduces the next-nearest-neighbor (plaquette diagonal) couping
with respect to the $\sqrt{N} \times \sqrt{N}$ network.

We apply this idea to the case of the $d=4$ {\em quantum} system.
To begin with, we set up the Hilbert-space bases 
$ \{ | \sigma_1, \sigma_2, \dots , \sigma_N \rangle \} $   ($\sigma_i=\pm1$)
for the quantum spins $\{ \sigma^\alpha_i \}$ ($i=1,2,\dots,N$).
These bases diagonalize the operator $\sigma^z_j$; namely,
\begin{equation}
\sigma^z_j | \{ \sigma_i \} \rangle = \sigma_j | \{ \sigma_i \} \rangle  ,
\end{equation}
holds.

We consider the one- and two-body interactions separately.
Namely,
we decompose the Hamiltonian (\ref{Hamiltonian}) into two sectors,
\begin{equation}
{\cal H} 
   = 
{\cal H}^{(2)} ( \{ J_i \} )
 + {\cal H}^{(1)} (\Gamma,H)
        .
\end{equation}
The component ${\cal H}^{(2)}$ originates from the spin-spin interaction,
which depends on the exchange couplings $\{ J_i \}$.
On the other hand,
the contribution ${\cal H}^{(1)}$ comes from the single-spin terms,
depending on the magnetic fields, $\Gamma$ and $H$.

First, we consider ${\cal H}^{(2)}$.
This component concerns the mutual connectivity among the $N$ spins,
and we apply Novotny's idea to represent the matrix elements.
We propose the following expression,
\begin{eqnarray}
\label{Hamiltonian2}
{\cal H}^{(2)}   & = & 
\frac{J_1}{2} \sum_{\tilde{\alpha}=\pm\alpha,\alpha \in A} H(\tilde{\alpha})
+\frac{J_2}{2} \sum_{\alpha , \beta \in A} 
       \sum_{\tilde{\alpha}=\pm \alpha}
       \sum_{\tilde{\beta}=\pm \beta}
   H(\tilde{\alpha}+\tilde{\beta})   \nonumber \\
& & +\frac{J_3}{2} \sum_{\alpha,\beta,\gamma \in A} 
 \sum_{\tilde{\alpha}=\pm \alpha}
 \sum_{\tilde{\beta}=\pm \beta}
  \sum_{\tilde{\gamma}=\pm\gamma}
   H(\tilde{\alpha}+\tilde{\beta}+\tilde{\gamma})  \nonumber \\
& & +\frac{J_4}{2} \sum_{\tilde{\alpha}=\pm 1}
                   \sum_{\tilde{\beta}=\pm N^{1/4}}
                   \sum_{\tilde{\gamma}=\pm N^{1/2}}
                   \sum_{\tilde{\delta}=\pm N^{3/4}}
   H(\tilde{\alpha}+\tilde{\beta}+\tilde{\gamma}+\tilde{\delta}) .
\end{eqnarray}
Here, the set $A$ consists of the elements, $A=\{ 1,N^{1/4},N^{1/2},N^{3/4} \}$.
The component $H(v)$ denotes
the $v$th-neighbor interaction for the $N$-spin alignment,
\begin{equation}
H_{ \{ \sigma_i \} ,\{ \tau_i \} }(v) = 
  \langle \{\sigma_i\}  | H(v) | \{ \tau_i\} \rangle  =
    \langle \{\sigma_i \} | T P^v | \{ \tau_i \} \rangle    ,
\end{equation}
with the exchange-interaction matrix,
\begin{equation}
\langle \{\sigma_i \} | T | \{ \tau_i \} \rangle = 
  -
\sum_{k=1}^{N} \sigma_k \tau_k 
   ,
\end{equation}
and
the translational operator $P$ satisfying,
\begin{equation}
P | \{ \sigma_i \} \rangle = | \{ \sigma_{i+1} \} \rangle  ,
\end{equation}
under the
periodic boundary condition.
The insertion of $P^v$ beside the $T$ operation is a key element to introduce the
coupling over the
$v$th-neighbor pairs.
The denominator $2$ in Eq. (\ref{Hamiltonian2}) 
compensates the duplicated sum.

Let us explain the meaning of the above formula, Eq. 
(\ref{Hamiltonian2}), 
more in detail.
As shown in Fig. \ref{figure1}, in the case of $d=2$, 
we made bridges over $N^{1/2}$th-neighbor pairs to lift up the
dimensionality to $d=2$ effectively.
In the case of $d=4$, by analogy,
we introduce the interaction distances such as $v=1,N^{1/4},N^{1/2}$ 
and $N^{3/4}$.
The first term in Eq. (\ref{Hamiltonian2}) thus represents the nearest-neighbor
interactions (with respect to the $d=4$-dimensional cluster).
Similarly, the remaining terms introduce the long-range interactions.
For example, the component $H(1+N^{1/4})$
introduces the next-nearest-neighbor (plaquette diagonal) interaction.
We emphasize that the idea of Novotny is
readily applicable to the quantum model.
[In short, our (quantum mechanical) formulation is additive.
On the contrary, Novotny's original formulation is multiplicative,
because his original formulation concerns the Boltzmann weight
rather than the Hamiltonian itself.]


Lastly, we consider the one-body part ${\cal H}^{(1)}$.
The matrix element is given by the formula,
\begin{equation}
\label{Hamiltonian1}
{\cal H}^{(1)}_{\{\sigma_i\}, \{\tau_i\}} = \langle \{\sigma_i \} | {\cal H}^{(1)}   | \{ \tau_i \} \rangle
   .
\end{equation}
The expression is quite standard,
because the component ${\cal H}^{(1)}$ simply
concerns the individual spins, and has nothing to do with
the connectivity among them.

The above formulas complete our basis to simulate the quantum Hamiltonian 
(\ref{Hamiltonian})
numerically.
In the next section, we perform the numerical simulation for $N=8,10, \dots , 28$.

\section{\label{section3}
Numerical results}

In Sec. \ref{section2}, we set up an explicit expression for
the Hamiltonian (\ref{Hamiltonian});
see Eqs. (\ref{Hamiltonian2}) and (\ref{Hamiltonian1}).
In this section, we diagonalize the Hamiltonian 
for $N=8,10,\dots ,28$
with the Lanczos algorithm.
We calculated
the first-excitation energy gap 
$\Delta E$ (rather than $\xi$).
The scaling relation for $\Delta E$ is given by,
\begin{equation}
\label{cylinder}
\Delta E =   L^{-\Omega} 
   g(\epsilon L^{y^*_t} , H L^{y^*_h} )   ,
\end{equation}
because $\Delta E \sim 1/\xi$ holds.
(As compared to Eq. (\ref{scaling_for_xi}),
our scaling relation is extended to include the magnetic field $H$
as well as the corresponding scaling index $y^*_h$.)
The reduced temperature $\epsilon$ is given by
 $\epsilon=\Gamma-\Gamma_c$ with the critical point $\Gamma_c$.
Note that the linear dimension $L$ satisfies
$L=N^{1/4}$, because the $N$ spins constitute the $d=4$-dimensional
cluster.

We fix the interaction parameters to,
\begin{equation}
\label{coupling_constants}
(J_1,J_2,J_3,J_4)=(1,0.15,0.05,0.05) ,
\end{equation}
and scan the transverse magnetic field $\Gamma$.
(We will also provide data for 
$(J_1,J_2,J_3,J_4)=(1,0,0,0)$ and
$(1,0.1,0.1,0.05)$ as a reference.)
The interaction parameters,
Eq. (\ref{coupling_constants}), are optimal in the sense
that the insystematic finite-size errors are 
suppressed satisfactorily.

\subsection{\label{section3_1}
Scaling behavior of Binder's cumulant ratio and the transition point}

Because
the scaling relation, Eq. (\ref{cylinder}), contains a number of free parameters,
it is ambiguous to determine these parameters simultaneously.
Actually,
in Ref. \cite{Novotny90,Novotny91}, the author fixed $\Omega=4/3$,
to determine the index $y^*_t$.

In this paper, we estimate the scaling indices independently
without resorting to any postulations.
For that purpose, we calculated an additional quantity, namely,
Binder's cumulant ratio \cite{Binder81},
\begin{equation}
\label{Binder_def}
U = 1- \frac{ \langle M^4 \rangle}
             {3 \langle M^2 \rangle^2 }  ,
\end{equation}
to determine the location of $\Gamma_c$.
Here, the brackets $\langle \dots \rangle$ denote the 
expectation value at the ground state.
The magnetic moment $M$ is given by $M=\sum_{i=1}^N \sigma^z_i$.
Because the cumulant ratio is dimensionless
($\Omega=0$), it obeys a simplified scaling relation;
\begin{equation}
\label{binder}
U= \tilde{U}(\epsilon L^{y^*_t} , H L^{y^*_h} )
     .
\end{equation}
Hence, the intersection point of the cumulant-ratio curves 
indicates
a location of
the critical point.
The scaling relation for the cumulant ratio, Eq. (\ref{binder}),
has been studied extensively for the $d=5$ hypercubic geometry
with the Monte Carlo method
\cite{Binder85,Binder85b,Parisi96,Jones05}.

In Fig. \ref{figure2},
we plotted the cumulant ratio for various $\Gamma$ and $N=8,10, \dots ,28$
with $H=0$ fixed; as mentioned above,
we fixed the exchange-coupling constants $\{ J_i \}$ to Eq. (\ref{coupling_constants}).
From the scale-invariant (intersection) point of the curves in Fig. \ref{figure2}, 
we observe a clear indication of criticality at $\Gamma_c \approx 12.5$.
In the subsequent analysis of Sec. \ref{section3_2}, we make use of
this information to determine the scaling indices.

This is a good position
to address why we fixed the exchange couplings to Eq. (\ref{coupling_constants}).
As a comparison,
we presented the cumulant ratio for various $\Gamma$
and $N=8,10,\dots,28$
in Fig. \ref{figure3}, where
we tentatively turn off the extended couplings
$J_{2,3,4}=0$.
Clearly, the data are scattered as compared to those of Fig. \ref{figure2}.
Such data scatter obscures the onset of the phase-transition point,
and prohibits detailed data analysis of criticality.
In order to improve the finite-size behavior, we surveyed
the parameter space $\{ J_i \}$, and found that 
the choice (\ref{coupling_constants}) is an optimal one.
Such elimination of finite-size errors has been utilized successfully in
recent numerical studies 
\cite{Blote96,Nishiyama06}.

\subsection{\label{section3_2}
Critical exponent $\Omega$}

Provided by the information on $\Gamma_c$ (Fig. \ref{figure2}),
we are able to determine the scaling indices from the scaling
relation (\ref{cylinder}).
In this section, we consider the index $\Omega$.

In Fig. \ref{figure4},
we plotted the approximate index,
\begin{eqnarray}
\label{indexI}
\Omega(L_1,L_2)=-
   \frac{
     \left.
     \ln ( \Delta E(N_1) / \Delta E(N_2) ) 
     \right|_{\Gamma=\Gamma_c(L_1,L_2)}   }{
      \ln (L_1 /L_2) }     ,
\end{eqnarray}
for $[2/(L_1+L_2)]^3$ with $8 \le N_1 < N_2 \le 28$;
note that $L_{1,2}=N_{1,2}^{1/4}$ holds.
The parameters are the same as those of Fig. \ref{figure2}.
The approximate transition point $\Gamma_c(L_1,L_2)$ 
is given by the
intersection point of the cumulant ratio for a pair of $(N_1,N_2)$; 
namely, it satisfies
\begin{equation}
U(N_1,\Gamma_c (L_1,L_2))=U(N_2,\Gamma_c (L_1,L_2))    .
\end{equation}
The least-squares fit to these data yields $\Omega=1.403(46)$ in the thermodynamic limit.
We carried out similar data analysis for $(J_2,J_3,J_4)=(0.1,0.1,0.05)$,
and obtained $\Omega=1.494(21)$.
As an error indicator, we accept the difference between them.
As a consequence we estimate the index as
\begin{equation}
\label{index1}
\Omega=1.40(15)      .
\end{equation}

Let us mention a number of remarks.
First, our result excludes the naively expected one $\Omega=1$.
Actually,
the result $\Omega > 1$ indicates that the correlation length
$L^\Omega$
develops more rapidly than the system size $L$ enlarges.
This feature may reflect the fact that the spin waves cost very
little energy large $L$.
Second, our result supports the generic formula $\Omega=4/3$,
Eq.
(\ref{ippan_cylinder}), advocated for the cylindrical geometry
in $d \ge 4$ \cite{Brezin82,Brezin85,Privman90}.
On the contrary,
it deviates from that of the hypercubic geometry (\ref{hypercubic});
we confirm this observation in the following sections.
Lastly, the validity of the abscissa scale (extrapolation scheme),
$1/L^3$,
in Fig. \ref{figure4} is not clear.
In Sec \ref{section3_4}, we inquire into the validity of the extrapolation scheme.

\subsection{\label{section3_3}
Critical exponents $y^*_t$ and $y^*_h$}

In Figs. \ref{figure5} and \ref{figure6}, we plotted the approximate indices,
\begin{equation}
\label{indexII}
-\Omega(L_1,L_2)+y^*_t(L_1,L_2)=
   \frac{
     \left.
     \ln ( \partial_\Gamma \Delta E(N_1) / \partial_\Gamma \Delta E(N_2) ) 
     \right|_{\Gamma=\Gamma_c(L_1,L_2)}   }{
      \ln (L_1 /L_2) }     ,
\end{equation}
and,
\begin{equation}
\label{indexIII}
-\Omega(L_1,L_2)+2y^*_h(L_1,L_2)=
   \frac{
     \left.
     \ln ( \partial^2_H \Delta E(N_1) / \partial^2_H \Delta E(N_2) ) 
     \right|_{H=0, \Gamma=\Gamma_c(L_1,L_2)}   }{
      \ln (L_1 /L_2) }     ,
\end{equation}
respectively,
for $[2/(L_1+L_2)]^3$ with $8 \le N_1 < N_2 \le 28$;
the parameters are the same as those of Fig. \ref{figure2}.
The least-squares fit to these data
yields the estimates, 
$-\Omega+y^*_t=1.396(21)$ and $-\Omega+2y^*_h=7.198(39)$,
in the thermodynamic limit.
Similarly for $(J_2,J_3,J_4)=(0.1,0.1,0.05)$, we obtained 
$-\Omega+y^*_t=1.359(12)$ and $-\Omega+2y^*_h=7.211(24)$.
Consequently, we estimate the scaling indices as
$-\Omega+y^*_t=1.4(1)$ and $-\Omega+2y^*_h=7.2(1)$.
Combining them with $\Omega=1.40(15)$, Eq. (\ref{index1}),
we arrive at,
\begin{equation}
\label{index2}
y^*_t=2.8(2)   ,
\end{equation}
and,
\begin{equation}
\label{index3}
y^*_h=4.3(1)   .
\end{equation}

Again,
our data exclude the naively expected values, $y^*_t=2$ and $y^*_h=3$.
Rather, our estimates are comparable with 
the generic formula, 
Eq. (\ref{ippan_cylinder});
actually,
the estimate $y^*_t=2.8(2)$ is quite consistent with the prediction
$y^*_t=8/3$, Eq. (\ref{ippan_cylinder}),
whereas the result $y^*_h=4.3(1)$ and the formula $y^*_h=4$,
Eq. (\ref{ippan_cylinder}),
are rather out of the error margin.
We attain more satisfactory agreement between the numerical result and
the formula by the data analysis
under the assumption $\Omega=4/3$ in the next section.
On the contrary,
our data conflict with the values, Eq. (\ref{hypercubic}),
anticipated for the hypercubic geometry.
Hence, the data suggest that the embedding geometry 
is influential on the finite-size scaling above the upper critical dimension.
We confirm this issue more in detail in the next section.

\subsection{\label{section3_4}
Scaling indices $y^*_t$ and $y^*_h$ under the assumption $\Omega=4/3$}

In Sec. \ref{section3_2}, we obtained an estimate $\Omega=1.40(15)$
being in good agreement with the
formula $\Omega=4/3$, Eq. (\ref{ippan_cylinder}).
In this section, we assume $\Omega=4/3$ \cite{Brezin82},
and estimate the remaining indices $y^*_t$ and $y^*_h$ 
under this hypothesis.

In Fig. \ref{figure7},
we plotted the approximate index,
\begin{equation}
\label{indexIV}
y^*_t (L_1,L_2)=
   \frac{
     \left.
     \ln ( \partial_\Gamma U(N_1) / \partial_\Gamma U(N_2) ) 
     \right|_{\Gamma=\tilde{\Gamma}_c(L_1,L_2)}   }{
      \ln (L_1 /L_2) }     ,
\end{equation}
for the abscissa scale $[2/(L_1 + L_2)]^3$ with $8 \le N_1 < N_2 \le 28$;
the parameters are the same as those of Figs. \ref{figure4}-\ref{figure6}.
Because we assumed $\Omega=4/3$ \cite{Brezin82}, we are able to determine the approximate
critical point $\tilde{\Gamma}_c(L_1,L_2)$ from the fixed point
of $L^{4/3} \Delta E(L)$; namely,
\begin{equation}
L_1^{4/3} \Delta E(N_1,\tilde{\Gamma}_c(L_1,L_2)) = 
L_2^{4/3} \Delta E(N_2,\tilde{\Gamma}_c(L_1,L_2))  .
\end{equation}
We notice that the data exhibit improved convergence to the thermodynamic limit.
The least-squares fit to these data yields $y^*_t=2.671(49)$
in the thermodynamic limit.
Similarly, we obtained $y^*_t=2.697(42)$ for 
$(J_2,J_3,J_4)=(0.1,0.1,0.05)$.
Consequently, we estimate,
\begin{equation}
y^*_t=2.67(10)
.
\end{equation}
This result is consistent with the above estimate $y^*_t=2.8(2)$, Eq. (\ref{index2}),
confirming the reliability of our analyses in Figs. \ref{figure4}-\ref{figure6}.
It is also in good agreement with the prediction 
$y^*_t=8/3$, Eq. (\ref{ippan_cylinder}).
On the contrary, our estimate excludes
the exponent $y^*_t=5/2$, Eq. (\ref{hypercubic}), advocated for the hypercubic geometry.

Similarly, in Fig. \ref{figure8}, we plotted the approximate 
index,
\begin{equation}
\label{indexV}
y^*_h (L_1,L_2)=  \frac{1}{2}
   \frac{
     \left.
     \ln ( \partial_H^2 U(N_1) / \partial_H^2 U(N_2) ) 
     \right|_{H=0,\Gamma=\tilde{\Gamma}_c(L_1,L_2)}   }{
      \ln (L_1 /L_2) }     ,
\end{equation}
for the abscissa scale $[2/(L_1 + L_2)]^3$ with $8 \le N_1 < N_2 \le 28$;
the parameters are the same as those of Figs. \ref{figure4}-\ref{figure6}.
The data exhibit an appreciable systematic finite-size deviation.
The least-squares fit to these data yields $y^*_h=4.021(60)$.
Similarly, we obtained $y^*_h=4.148(36)$ for 
$(J_2,J_3,J_4)=(0.1,0.1,0.05)$.
Consequently, we estimate,
\begin{equation}
y^*_h=4.0(2)
.
\end{equation}
Again, 
the result is quite consistent with the prediction 
$y^*_h=4$, Eq. (\ref{ippan_cylinder}).
In other worlds, this agreement 
indicates that the extrapolation scheme (abscissa scale) $1/L^3$ is sensible.


Let us mention a few remarks.
First, the data in Figs. \ref{figure7} and \ref{figure8} exhibit suppressed finite-size corrections
owing to the assumption $\Omega=4/3$.
This feature was observed in Refs. \cite{Novotny90,Novotny91},
where the author estimated $y^*_t$ reliably under $\Omega=(d-1)/3$;
see the Introduction.
Our analysis shows that the assumption yields a reliable estimate for $y^*_h$ 
as well as $y^*_t$.
Second, our data confirm the self-consistency of our analyses
performed in Figs. \ref{figure4}-\ref{figure8}.
Particularly, the data justify
the extrapolation scheme
with the abscissa scale $1/L^3$.
As a matter of fact,
in Ref. \cite{Luijten96}, the authors observed 
notable finite-size corrections to the cumulant ratio $U$
obeying $L^{d- 2 y^*_h}$.
In our case ($d=5$ cylindrical geometry), the power 
should read $d-2y_h^*=-3$.
Hence, our numerical data support their claim.
Lastly,
as to the convergence of $U$ to the thermodynamic limit,
there arose controversies
\cite{Chen98,Chen98b,Chen00,Luijten99,Binder00,Mon96,Luijten97,Mon97,Blote97};
it has been reported that there appear
unclarified finite-size corrections to $U$, which
prohibit us to take reliable extrapolation to the thermodynamic limit.
In this paper, we avoided the subtlety by eliminating
finite-size errors with
the (finitely-tuned) extended interactions;
see Figs. \ref{figure2} and \ref{figure3}.
We consider that such a technique would be significant
for the study of high-dimensional systems,
where the available system size is restricted.

\section{Summary and discussions}

We studied the finite-size-scaling behavior of the $d=5$ Ising model embedded 
in the cylindrical geometry.
Our aim is to see an influence of the embedding geometry (boundary condition)
on the scaling relation, Eq. (\ref{cylinder});
the embedding geometry should alter the scaling indices $\Omega$ and $y^*_{t,h}$
above $d>4$ \cite{Brezin82,Brezin85,Privman90}.
For that purpose, we employed the transfer-matrix method (Sec. \ref{section2}),
and implemented Novotny's technique 
\cite{Novotny90} to treat a variety of system sizes
$N=8,10, \dots ,28$.
Moreover, we made use of an
equivalence between the $d=5$ (classical) Ising model and its $d=4$ quantum counterpart;
the latter version is computationally less demanding with the universality class retained.

We analyzed the simulation data with the finite-size scaling relation, Eq. (\ref{cylinder}),
and obtained the scaling indices as 
$\Omega=1.40(15)$, $y^*_t=2.8(2)$, and $y^*_h=4.3(1)$.
Additionally, under $\Omega=4/3$, we estimate
$y^*_t=2.67(10)$ and $y^*_h=4.0(2)$.
The indices exclude the naively expected ones,
$\Omega=1$, $y^*_t=2$, and $y^*_h=3$, reflecting
the violation of hyperscaling in large dimensions.
Clearly, our data support the generic formulas, Eq. (\ref{ippan_cylinder}),
advocated for the cylindrical geometry in $d \ge 4$ \cite{Brezin82,Brezin85,Privman90}.
On the contrary, our data conflict with the values for the hypercubic geometry,
Eq. (\ref{hypercubic}).    
Our result demonstrates that
the embedding geometry is indeed influential on 
the scaling indices.

\begin{figure}
\includegraphics[width=100mm]{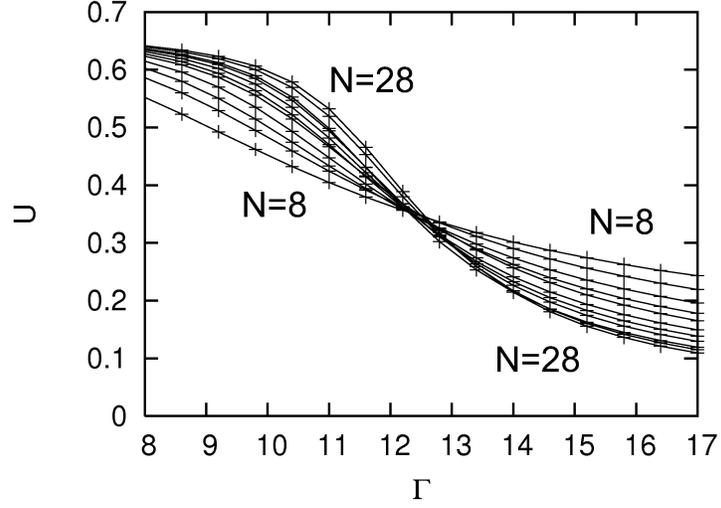}%
\caption{
\label{figure2}
Binder's cumulant ratio $U$ 
(\ref{Binder_def}) is plotted for the transverse magnetic field $\Gamma$
and the system sizes 
$N=8,10, \dots ,28$
with the fixed exchange couplings, Eq. (\ref{coupling_constants}).
We observe a clear indication of criticality at $\Gamma_c \approx 12.5$.
Apparently, the finite-size-scaling behavior is improved as compared to that of 
Fig. \ref{figure3}, where we turned off
the extended interactions $J_{2,3,4}=0$.
}
\end{figure}

\begin{figure}
\includegraphics[width=100mm]{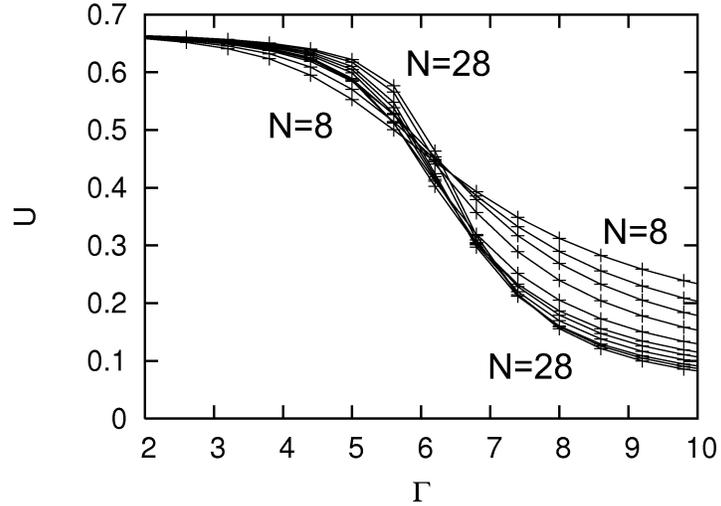}%
\caption{
\label{figure3}
Tentatively, we turned off the extended interactions
($J_{2,3,4}=0$),
and calculated the cumulant ratio $U$ 
(\ref{Binder_def}) for various $\Gamma$ and 
$N=8,10, \dots ,28$.
We notice that the data are scattered as compared to those in Fig. \ref{figure2}.
}
\end{figure}

\begin{figure}
\includegraphics[width=100mm]{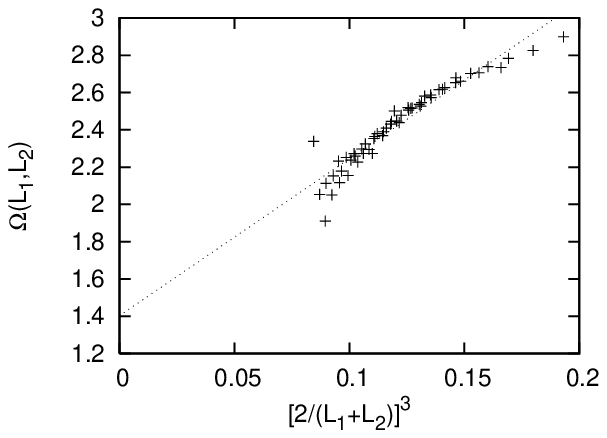}%
\caption{
\label{figure4}
The approximate critical index $\Omega(L_1,L_2)$ 
(\ref{indexI}) is plotted for $[2/(L_1+L_2)]^3$ 
with $8 \le N_1<N_2 \le 28$ ($L_{1,2}=N_{1,2}^{1/4}$);
the parameters are the same as those of Fig. \ref{figure2}.
The least-squares fit to these data yields 
$\Omega=1.403(46)$
in the thermodynamic limit $L \to \infty$.
}
\end{figure}

\begin{figure}
\includegraphics[width=100mm]{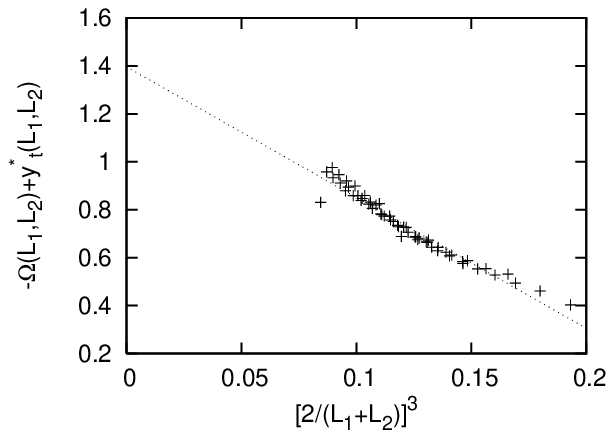}%
\caption{
\label{figure5}
The approximate critical index $-\Omega(L_1,L_2)+y^*_t(L_1,L_2)$ 
(\ref{indexII}) is plotted for $[2/(L_1+L_2)]^3$
with $8 \le N_1<N_2 \le 28$ ($L_{1,2}=N_{1,2}^{1/4}$);
the parameters are the same as those of Fig. \ref{figure2}.
The least-squares fit to these data yields 
$-\Omega+y^*_t=1.396(21)$
in the thermodynamic limit $L \to \infty$.
}
\end{figure}

\begin{figure}
\includegraphics[width=100mm]{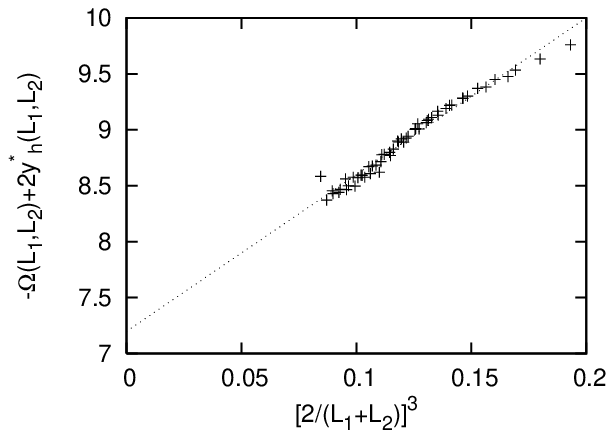}%
\caption{
\label{figure6}
The approximate critical index $-\Omega(L_1,L_2)+2y^*_h(L_1,L_2)$ 
(\ref{indexIII}) is plotted for $[2/(L_1+L_2)]^3$
with $8 \le N_1<N_2 \le 28$ ($L_{1,2}=N_{1,2}^{1/4}$);
the parameters are the same as those of Fig. \ref{figure2}.
The least-squares fit to these data yields 
$-\Omega+2y^*_h=7.198(39)$
in the thermodynamic limit $L \to \infty$.
}
\end{figure}

\begin{figure}
\includegraphics[width=100mm]{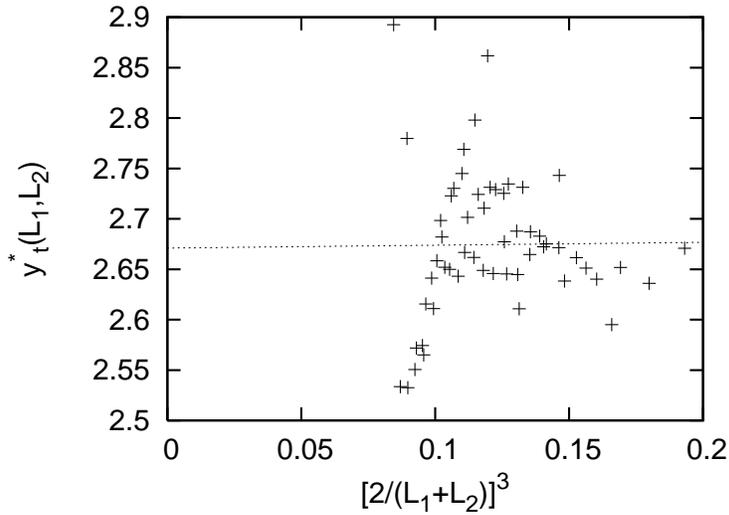}%
\caption{
\label{figure7}
The approximate critical index $y^*_t(L_1,L_2)$ (\ref{indexIV}) 
is plotted for $[2/(L_1+L_2)]^3$
with $8 \le N_1<N_2 \le 28$ ($L_{1,2}=N_{1,2}^{1/4}$);
the parameters are the same as those of Fig. \ref{figure2}.
The least-squares fit to these data yields $y^*_t=2.671(49)$
in the thermodynamic limit $L \to \infty$.
The result is consistent with Eq. (\ref{index2}),
confirming the reliability of our analysis.
}
\end{figure}

\begin{figure}
\includegraphics[width=100mm]{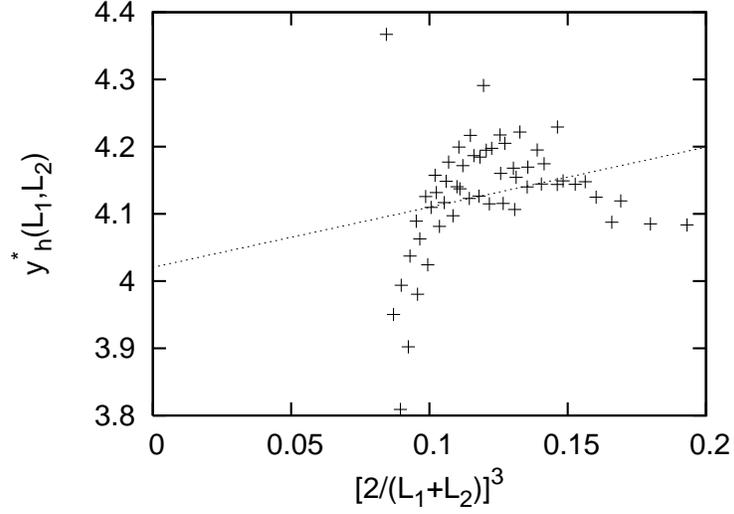}%
\caption{
\label{figure8}
The approximate critical index $y^*_h(L_1,L_2)$ (\ref{indexV}) 
is plotted for $[2/(L_1+L_2)]^3$
with $8 \le N_1<N_2 \le 28$ ($L_{1,2}=N_{1,2}^{1/4}$);
the parameters are the same as those of Fig. \ref{figure2}.
The least-squares fit to these data yields $y^*_h=4.021(60)$
in the thermodynamic limit $L \to \infty$.
The result is consistent with the prediction
(\ref{ippan_cylinder}).
}
\end{figure}

Lastly, let us mention a few remarks.
First, we stress that 
the violation of hyperscaling above the upper critical dimension
$\tilde{d}$
is not necessarily an issue of pure academic interest.
For example, a class of long-range interaction 
\cite{Luijten96}
suppresses the upper critical dimension 
to an experimentally accessible regime $\tilde{d}<3$.
Second, 
there arose controversies 
\cite{Chen98,Chen98b,Chen00,Luijten99,Binder00,Mon96,Luijten97,Mon97,Blote97}
concerning
the subdominant finite-size effect (corrections to scaling) above $\tilde{d}$.
More specifically, 
the Binder-cumulant data
exhibit unexpectedly slow convergence to
the thermodynamic limit.
In this paper, we avoided this subtlety
by extending (tuning) the exchange-coupling constants 
to Eq. (\ref{coupling_constants}), where we observed eliminated
finite-size errors.
Actually, from Figs. \ref{figure2} and \ref{figure3}, 
we notice that the elimination was successful.
Our data indicate that the (dominant) finite-size errors obey
the power law $L^{-3}$ as claimed in Ref. \cite{Luijten96}.
We consider that the elimination of finite-size errors
is significant for the study of high-dimensional systems,
where the available system size is restricted severely.

\begin{acknowledgments}
This work is supported by a Grant-in-Aid 
(No. 18740234) from Monbu-Kagakusho, Japan.
\end{acknowledgments}


\end{document}